# Development and pilot evaluation of a virtual reality simulator for HDR prostate brachytherapy

Anton Varlukhin[1], Mackenzie Smith[2], Fahad Alam[3], Amandeep Tagger[2,4], Gerard Morton[2,4], Moti Paudel[2,4], Andrew Loblaw[2,4], Lucas Mendez[5,6], Douglas Hoover[5,6], Raffi Karshafian[1], Humza Nusrat[7,8,1]

(1) Toronto Metropolitan University, Toronto, Canada, (2) Sunnybrook Health Sciences Center, Toronto, Canada, (3) Department of Anesthesiology, Temerty Medicine, University of Toronto, Toronto, Canada, (4) Department of Radiation Oncology, Temerty Medicine, University of Toronto, Toronto, Canada, (5) Verspeeten Family Cancer Center, London Health Sciences Center, London, Canada, (6) Department of Radiation Oncology, Western University, London, Canada, (7) Henry Ford Health, Detroit, USA, (8) Department of Radiation Oncology, Michigan State University, East Lansing, USA

Correspondence: Humza Nusrat, PhD, nusrathu@msu.edu.

# 1. Abstract

Purpose: To develop a virtual reality simulator for high dose rate prostate brachytherapy and to test whether participation is associated with immediate gains in self-reported confidence across predefined procedural domains in two cohorts.

Methods: Two modules were developed and implemented using Unreal Engine: patient preparation and template guided needle insertion. Oncology staff and trainees completed pre and post surveys that assessed confidence for recalling steps, explaining steps, identifying equipment, and explaining equipment function. Studies were conducted at the Hands On Brachytherapy Workshop (HOWBT) in London, Ontario, and at Sunnybrook Odette Cancer Centre in Toronto, Ontario. Paired Wilcoxon signed rank tests with two-sided p values compared before and after scores within each module.

Results: Patient preparation (N=11) confidence increased for recalling steps (W=65, p=0.002), explaining steps (W=51, p = 0.023), identifying equipment (W=65, p=0.002), and explaining equipment function (W=60, p=0.0078). Needle insertion (N=27) confidence increased for recalling steps (W=292, p<0.001), explaining steps (W=347, p<0.001), identifying equipment (W=355, p<0.001), and explaining equipment function (W=354, p<0.001).

Conclusion: The simulator was feasible to deploy and was associated with higher self-reported confidence across key domains immediately after training. Findings may inform future curriculum design and implementation work.

# 2. Introduction

Within North America, clinical use of brachytherapy has declined relative to external beam radiation therapy modalities, despite strong local control for selected genitourinary and gynecologic cancer sites [1], [2]. Training for high dose rate (HDR) prostate brachytherapy requires coordinated effort from radiation oncologists, medical physicists, therapists, nurses, and anesthesia teams. Opportunities to learn in clinic are constrained by case volume, operating room time, and space in academic centers, which reduces procedural exposure and erodes trainee confidence [3], [4].

Competency pathways exist, including American Brachytherapy Society fellowships and Canadian certification programs, but coverage and evaluation remain uneven. Surveys report that many practitioners receive no training beyond residency, some discontinue prostate brachytherapy due to infrastructure and coordination barriers, and a substantial fraction of programs lack practical skills assessment beyond written and oral examinations [4-7]. These gaps motivate efficient and scalable educational tools that supplement clinical teaching without competing for operating room resources.

Virtual reality (VR) offers an immersive supplement to traditional instruction. In radiation oncology, simulation platforms such as the Virtual Environment for Radiotherapy Training (VERT) have supported learning around linear accelerator workflows, and gynecologic brachytherapy training delivered with 360-degree video have improved trainee confidence and familiarity with equipment [8-11]. The recent availability of commodity head mounted displays and real time engines enables creation of realistic procedural environments at relatively low marginal cost. Across these applications, the most consistent benefit reported is higher self-reported confidence and engagement, with mixed evidence for transfer of skill depending on context.

Access constraints are amplified in low- and middle- human development indexed (HDI) countries, where radiotherapy services remain unevenly distributed and travel burden and wait times contribute to treatment abandonment [12-14]. While the present study does not evaluate implementation or cost, portable simulation has potential to support education where clinical throughput is limited and can therefore be considered a motivating context for future work.

We developed a virtual reality simulator for high dose rate prostate brachytherapy composed of two modules: patient preparation and template guided needle insertion. We evaluated immediate changes in self-reported confidence across four predefined domains and compared before and after scores within each module using paired Wilcoxon signed rank tests.

We hypothesized that participation in the simulator would be associated with higher median self-reported confidence after the session compared with before the session for recalling procedural steps as the primary outcome, with parallel increases for explaining steps, identifying equipment, and explaining equipment function as secondary outcomes. Consistent with the Kirkpatrick framework, this study assesses level one reaction and self-efficacy and does not claim knowledge, technical skill, or clinical performance.

## 3. Methods

### 3.1. Setting and participants

We conducted two single group pre and post evaluations: a patient preparation module at the Hands-on Brachytherapy (HOWBT) Workshop in London, Ontario, and a template guided needle insertion module at the Sunnybrook Odette Cancer Centre in Toronto, Ontario. Participants were oncology staff and trainees, including radiation oncologists, medical physicists, and radiation therapists. Only participants who completed both the pre survey and the post survey were included in paired analyses. All surveys were anonymous and collected without patient data.

### 3.2. Simulator

Both modules were built in Unreal Engine with room and equipment assets modeled in Blender. Reference photographs and room measurements from the Sunnybrook brachytherapy suite were used with fSpy to align images and dimensions for accurate digitization. The simulator ran on an Intel Core i7 12700F with 16 GB of RAM and an Nvidia RTX 3070. A Meta Quest Pro head mounted display provided optical hand tracking and approximately 3.5 MP per eye. Figure 1 shows the modeled operating room with the ultrasound cart and probe, catheter needles, patient on the table, and the anesthesia machine. Software versions were those current to development. Unreal Engine and Blender as of July 2024, and GraphPad Prism as of December 2024. Third party assets included the anesthesia machine and ultrasound cart from a commercial asset library, and the patient anatomy derived from an open-source male model containing rectum, bladder, urethra, seminal vesicles, and prostate.  Ultrasound displays were generated by virtual cameras parented to the probe and therefore did not model acoustic propagation, beamforming, attenuation, or speckle; images should be interpreted as view proxies rather than physics-based ultrasound.

### 3.3. Tasks

In the patient preparation module, participants identified required equipment and ordered key preparation steps for an HDR prostate case. In the needle insertion module, participants positioned the transrectal ultrasound (TRUS), interacted with the template coordinate interface, and steered virtual catheters toward target locations in the prostate model. Equipment interaction used Unreal Engine's node-based logic with grabbing mechanics and a template selection interface. Catheters changed color on contact with anatomy: red in organs at risk (OARs) and green in prostate (tumor target). Axial and sagittal B-mode ultrasound views were generated by virtual cameras attached to the TRUS and displayed on the ultrasound cart monitor. Precise needle placement was not simulated with full physical fidelity; a visual steering guide and contact color change operated as augmented feedback features to support learning. Figure 2 illustrates the needle insertion module with the TRUS in the rectum and a catheter in the prostate.

### 3.4. Survey instrument and outcomes

Pre and post surveys were administered using Google Forms. Items captured role, prior HDR prostate brachytherapy exposure, and prior VR experience. Four confidence constructs were assessed on a five-point Likert scale with anchors of one (equal to no prior experience) and five (equal to well versed): recalling steps, explaining steps, identifying equipment, and explaining equipment function. The primary outcome was the change in median self-reported confidence for recalling steps. Secondary outcomes were parallel changes for explaining steps, identifying equipment, and explaining equipment function. Post surveys also included free text feedback and a checklist to record symptoms such as nausea, dizziness, or headache. This study assesses Kirkpatrick level one reaction and self-efficacy and does not measure knowledge or technical skill.

### 3.5. Statistical analysis

Because outcomes were ordinal and paired, before and after scores were compared within each module using the paired Wilcoxon signed rank test with two-sided p values. Differences of zero were excluded from the rank sum as per the standard Wilcoxon matched pairs procedure, and tied absolute differences were assigned average ranks. Exact two-sided p values are reported. We report the number of paired observations, the Wilcoxon W statistic, and exact p values. Analyses were conducted in GraphPad Prism. The significance threshold alpha was 0.05 two sided. Given the pilot design, no adjustment for multiple comparisons was applied. Only complete pre and post pairs were analyzed.

### 3.6. Qualitative Analysis

Free text comments from post session surveys were exported to a spreadsheet without identifiers. One rater with domain familiarity performed an inductive descriptive content review. Short phrases were coded to provisional labels, which were then grouped into broader themes. No second rater was used and inter-rater agreement was not calculated. Results are presented as counts of comments within themes to summarize areas for improvement and future development.

### 3.7. Ethics

Institutional processes were followed for both evaluations. The template guided needle insertion evaluation was conducted as a quality improvement activity at Sunnybrook Odette Cancer Centre. The patient preparation evaluation was reviewed and approved by the Toronto Metropolitan University Research Ethics Board under protocol number 2024-354. Participation was voluntary. Surveys were anonymous and contained no patient data. Participants provided informed consent before beginning the survey and were informed that they could stop at any point.

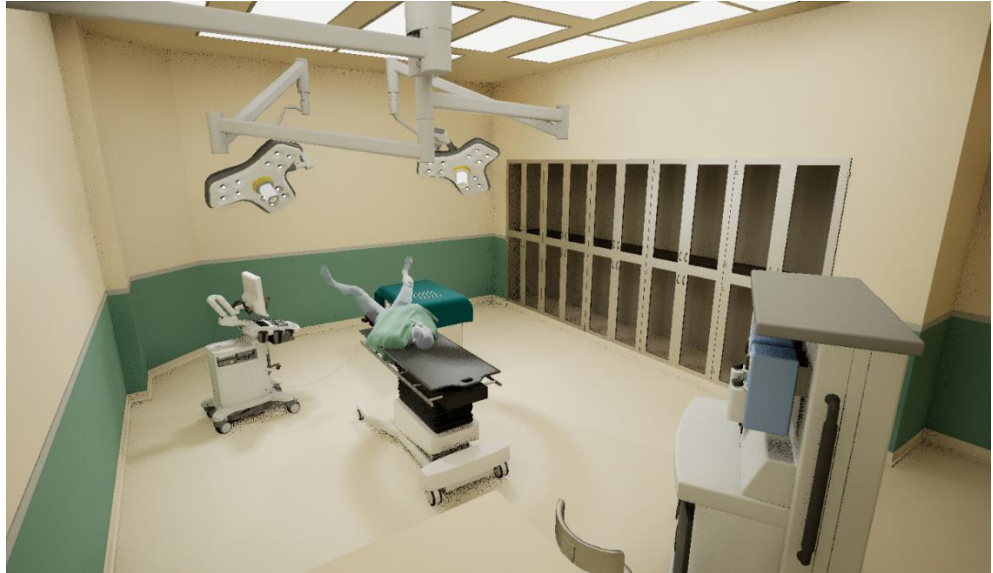

**Figure 1:** Modeled brachytherapy operating room environment with ultrasound cart and probe, catheter needles, patient on table, and anesthesia machine. This image documents the simulator layout used in both modules.

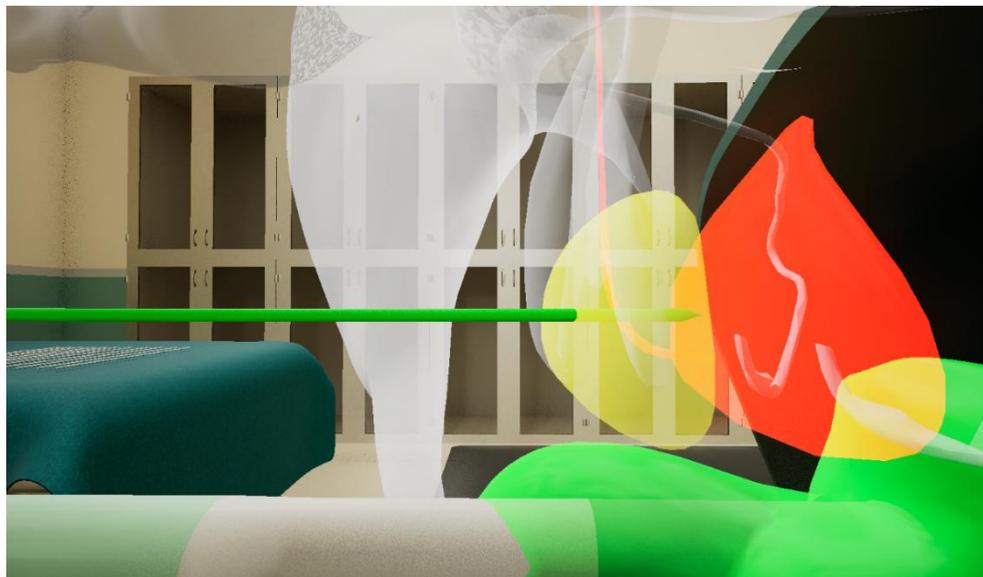

**Figure 2**: Needle insertion module screenshot showing a catheter in the prostate and the TRUS in the rectum, with contact color coding and the guidance overlay active. This image illustrates core interactions evaluated in the needle module.

# 4. Results

Across the patient preparation cohort of eleven participants and the needle insertion cohort of twenty-seven participants, median self-reported confidence increased from before to after for recalling steps, explaining steps, identifying equipment, and explaining equipment function. Within each module, paired Wilcoxon tests were statistically significant for all four constructs.

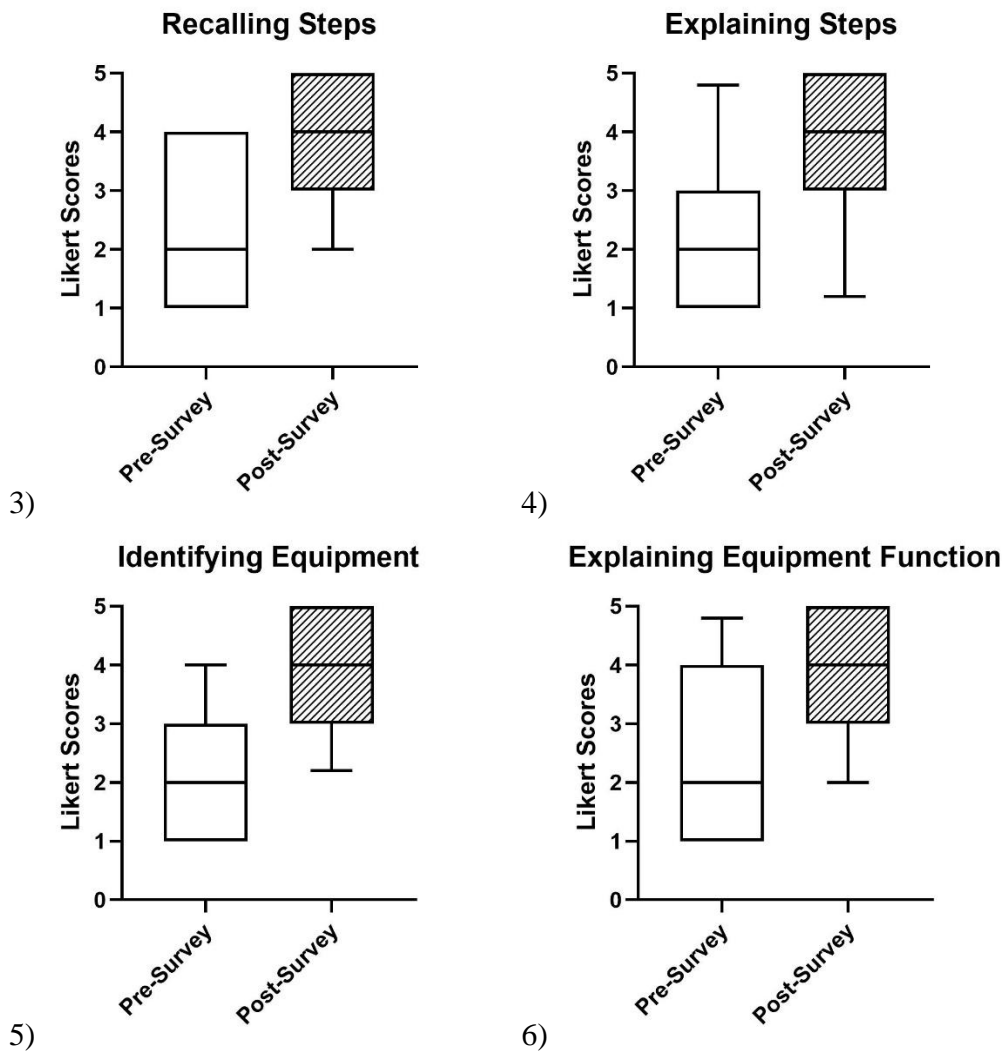

3)    4)

5)    6)

**Figure 3** (top-left): Median Likert scores related to the participant's ability to recall steps associated with patient preparation procedures pre / post. **Figure 4** (top-right):

Median Likert score values related to the participant's ability to explain steps associated with patient preparation procedures pre / post. **Figure 5** (bottom-left): Median Likert score values related to the participant's ability to identify equipment associated with patient preparation procedures pre / post. **Figure 6** (bottom-right): median Likert score values related to the participant's ability to explain equipment function associated with patient preparation procedures pre / post.

4.1. **Patient preparation module**

Eleven participants completed paired surveys at the Hands On Brachytherapy Workshop in London, Ontario. Two of eleven reported prior virtual reality exposure. Three of eleven reported symptoms during the session one nausea, one dizziness, one headache and all completed the survey.

Recalling steps had increased after the training session ($W = 65$, $p = 0.002$). The explanation of steps increased after the training session ($W = 51$, $p = 0.023$). Identification of equipment increased after the training session ($W = 65$, $p = 0.002$). Explanation of equipment function increased after the training session ($W = 60$, $p = 0.0078$).

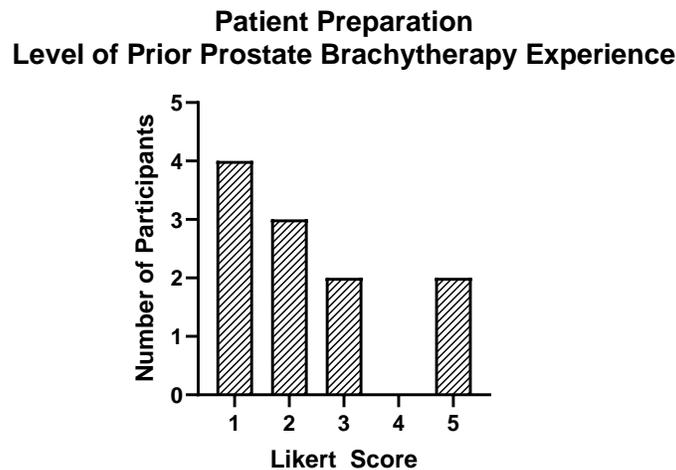

**Figure 7:** Frequency distribution histogram of self-reported levels of prior prostate brachytherapy experience. Median Likert score value was 2.

The median prior HDR prostate brachytherapy experience score was two, indicating limited prior exposure for most participants.

4.2. **Needle insertion module**

Twenty-seven participants completed paired surveys at the Sunnybrook Odette Cancer Centre. Thirteen of twenty-seven reported prior virtual reality exposure, generally infrequent. Four of twenty-seven reported symptoms during the session dizziness, discomfort, or nausea and all completed the survey.

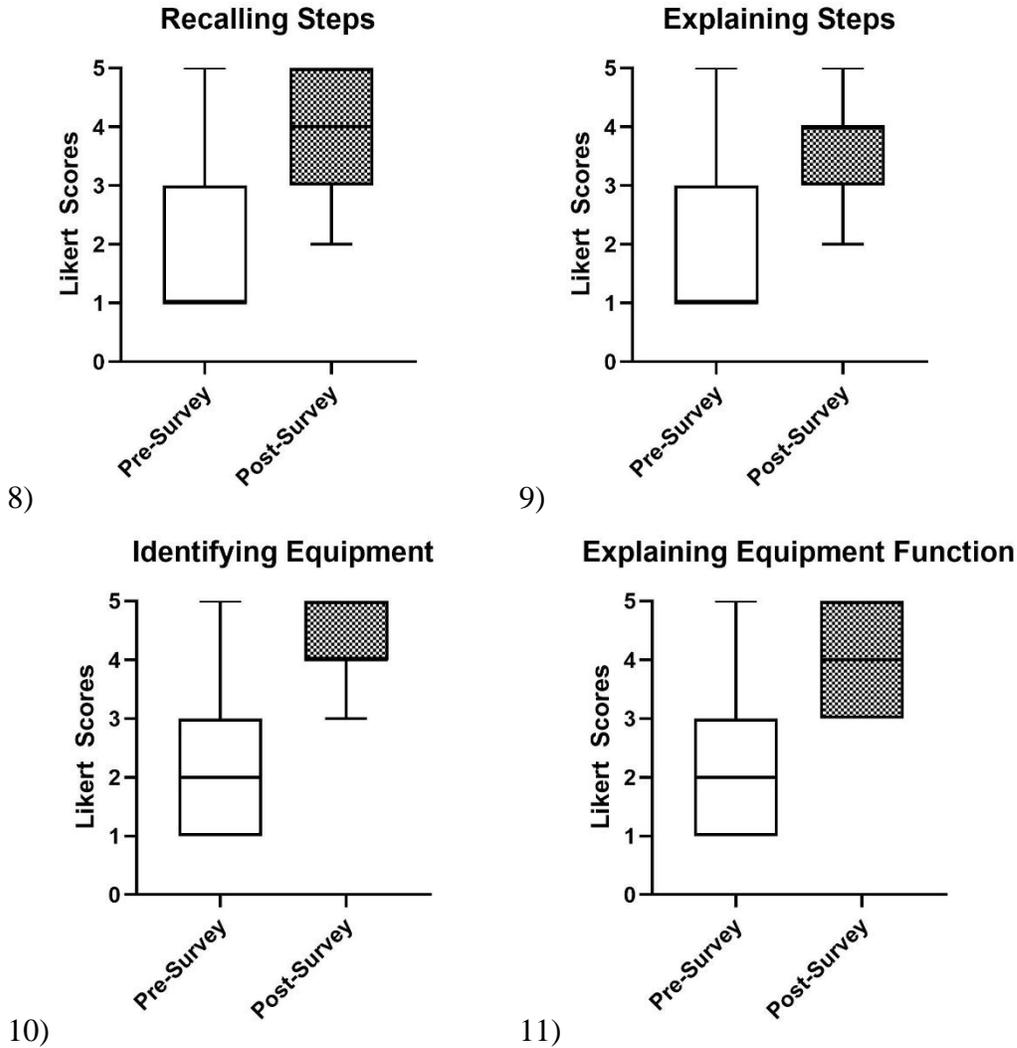

8)   9)   10)   11)

**Figure 8** (top-left): Median Likert score values related to the participant's ability to recall steps of associated with HDR procedures pre / post. **Figure 9** (top-right): Median Likert score values related to the participant's ability to explain steps associated with HDR procedures pre / post. **Figure 10** (bottom-left): Median Likert score values related to the participant's ability to identify equipment associated with HDR procedures pre / post. **Figure 11** (bottom-right): median Likert score values related to the participant's ability to explain equipment function associated with HDR procedures pre / post.

Recalling steps had increased after the training session (W = 292, p < 0.001). The explanation of steps increased after the training session (W = 347, p < 0.001). Identification of equipment increased after the training session (W = 355, p < 0.001). Explanation of equipment function increased after the training session (W = 354, p < 0.001).

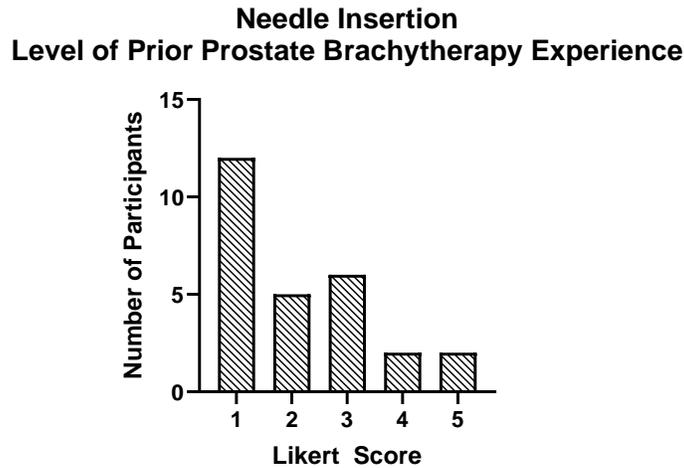

**Figure 12:** Frequency distribution histogram of self-reported levels of prior prostate brachytherapy experience. Median Likert score value was 2.

The median prior HDR prostate brachytherapy experience score was two, again indicating limited prior exposure for most participants.

4.3. **Feedback from free text responses**

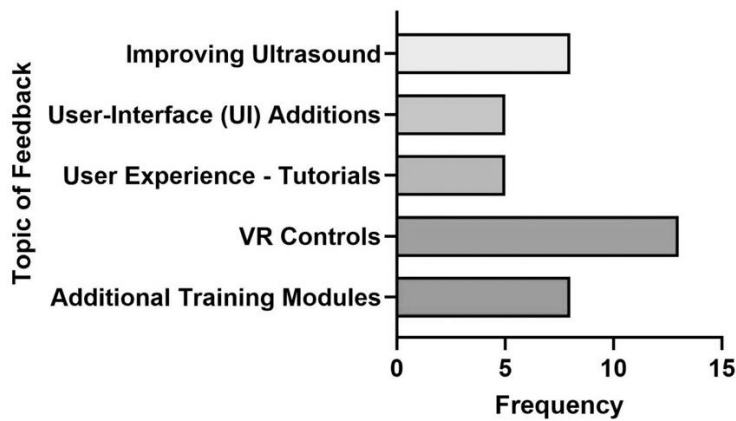

**Figure 13:** Frequency of themes from free text feedback.

Participants most frequently commented on the control scheme, reporting that controls were not intuitive. Additional themes included requests for more realistic ultrasound images, the ability to rotate the probe using the stepper, a movable ultrasound screen, a template grid overlay, user interface refinements for template coordinate selection, and a brief tutorial at the start of the session.

### 4.4. **Summary table of paired tests**

Table 1 summarizes the paired Wilcoxon results for each construct.

**Table 1. Paired Wilcoxon results for self-reported confidence by module and construct**
All tests are paired Wilcoxon signed rank, two sided. N is the number of paired observations.

| *Module* | *Construct* | N | W | p |
| --- | --- | --- | --- | --- |
| *Patient preparation* | Recalling steps | 11 | 65 | p = 0.002 |
| *Patient preparation* | Explaining steps | 11 | 51 | p = 0.023 |
| *Patient preparation* | Identifying equipment | 11 | 65 | p = 0.002 |
| *Patient preparation* | Explaining equipment function | 11 | 60 | p = 0.0078 |
| *Needle insertion* | Recalling steps | 27 | 292 | p < 0.001 |
| *Needle insertion* | Explaining steps | 27 | 347 | p < 0.001 |
| *Needle insertion* | Identifying equipment | 27 | 355 | p < 0.001 |
| *Needle insertion* | Explaining equipment function | 27 | 354 | p < 0.001 |

# 5. Discussion

This pilot evaluated a virtual reality training approach for HDR prostate brachytherapy and examined whether participation was associated with immediate gains in self-reported confidence across four domains in oncology staff and trainees. Within each module, paired Wilcoxon tests were significant for recalling steps, explaining steps, identifying equipment, and explaining equipment function. These findings are consistent across the patient preparation cohort and the needle insertion cohort and align with our prespecified hypothesis.

The simulator was feasible to deploy in two settings and included interactive room and equipment models, transrectal ultrasound views, and augmented feedback features such as contact color change and a visual guidance overlay that supported the training workflow. Participants also provided structured free text feedback that pointed to opportunities to improve the control scheme, ultrasound realism, a movable display, user interface refinements for template coordinate selection, and a brief tutorial. Consistent with the Kirkpatrick framework, the outcomes represent level one reaction and self-efficacy rather than knowledge or technical skill.

Virtual reality training has been reported to reduce the time required to complete training and operative tasks in other procedural domains. In a meta-analysis of laparoscopic training, Nagendran and colleagues compared virtual reality supplementation with no supplemental training and with box trainer approaches [11]. Across three trials totaling forty-nine participants, virtual reality reduced operative time by a mean of 11.76 minutes versus no supplemental training $p < 0.0001$. In two studies totaling thirty-three residents, objective performance metrics were significantly better after virtual reality training than with no supplemental training $p = 0.00049$. In a single trial with nineteen residents that compared virtual reality with a box trainer, operative performance favored virtual reality, although confidence outcomes were not reported [11].

Within radiation oncology, immersive platforms such as the Virtual Environment for Radiotherapy Training provide life sized linear accelerator and CT models that supplement conventional teaching [8]. In a randomized trial in an undergraduate radiography course, Ketterer et al reported that students who used VERT felt more confident before clinical placements, but assessment grades did not differ from conventional teaching [15]. Taken together, these studies suggest that simulation can improve learner confidence and engagement, while effects on objective performance depend on context and outcome measured.

Our evaluation focused on self-reported confidence after participation in a virtual reality simulator for HDR prostate brachytherapy. The study did not include a control arm and did not assess knowledge, technical skill, or clinical performance. Within that scope, pre and post comparisons support an association between simulator participation and higher confidence across predefined domains.

Related work in brachytherapy shows a similar pattern at the level of perceived confidence. Taunk and colleagues used a head mounted display with a 360-degree video for gynecologic HDR instruction and observed increases in confidence for assembling applicators, insertion, and comfort performing the procedure p < 0.001 [9]. Our prostate focused findings are directionally consistent with those results, while differing in clinical context and modality.

Strengths include two independent implementations across two locations, a clearly defined simulator workflow that mapped to patient preparation and template guided needle insertion, and consistent within person changes across all four confidence domains. The use of a single analysis approach with paired Wilcoxon tests and two-sided p values avoids nonstandard metrics and keeps inference aligned with the study aim.

Limitations are inherent to the design. The project measured immediate self-reported confidence only. Knowledge, technical skill, retention, transfer to clinic, and patient outcomes were not assessed. The evaluations were single session without a control group and used small convenience samples, which limited generalizability and made results susceptible to novelty and social desirability effects. A subset of participants reported symptoms such as nausea, dizziness, or discomfort, and the physical spaces used for testing constrained movement, which may have influenced user experience. The simulator incorporated augmented feedback guides and contact color changes that support learning but do not represent full physical fidelity, which is appropriate for early training but should be interpreted when considering external validity.

Sex or gender were not collected, and subgroup analyses were not possible. Prior work in simulation-based education suggests that learner characteristics can interact with perceived confidence and performance, but the direction and magnitude of such effects vary by context [16] [17]. Future evaluations should prospectively capture sex or gender along with prior experience to assess whether confidence changes differ across subgroups. The present findings should therefore be interpreted as averages across a heterogeneous cohort.

Portable simulation may be a useful supplementary tool where case volume or access to procedural environments is limited. These findings support the feasibility of a virtual environment to raise self-reported confidence for key procedural domains in HDR prostate brachytherapy. We did not evaluate cost or implementation, and feasibility in low- and middle-HDI settings remains to be determined, but access constraints in those settings provide a motivating context for future work.

# 6. Conclusion

This pilot provides an early foundation for virtual reality-based education in HDR prostate brachytherapy. The simulator was feasible to deploy at two locations and was associated with higher self-reported confidence across four procedural domains in oncology staff and trainees. The outcomes reported here reflect Kirkpatrick level one reaction and self-efficacy. We did not measure knowledge, technical skill, retention, transfer to clinic, or patient outcomes, and the single session design with small convenience samples limits generalizability. Taken together, these findings support virtual reality as a supplementary tool for curriculum planning in HDR prostate brachytherapy, particularly when case volume or access to procedural environments is constrained. Cost and implementation were not evaluated, including in low- and middle- HDI settings, and should be addressed in subsequent evaluations.

# 7. Data Availability

Deidentified survey data for all pre and post items, the survey instruments, and representative screenshots of the simulator workflow will be made available on request to the corresponding author. Third party 3D assets are subject to vendor licenses and cannot be redistributed. Project specific logic for the template interface and catheter color change can be shared as pseudocode on request.